\documentclass[useAMS,usenatbib]{mnras}
\usepackage[dvips]{graphicx}
\usepackage[english]{babel}
\usepackage{amsmath}
\usepackage{amssymb}
\usepackage{textcomp}
\usepackage{natbib}

\title[The environmental dependence of gas accretion]{The environmental dependence of gas accretion onto galaxies: quenching satellites through starvation}
\author[F. van de Voort et al.]{Freeke~van~de~Voort$^{1,2,3,4}$\thanks{E-mail: freeke.vandevoort@h-its.org}, Yannick~M.~Bah\'e$^{5}$, Richard~G.~Bower$^{6}$, 
\newauthor
Camila~A.~Correa$^{7}$, Robert~A.~Crain$^{8}$, Joop~Schaye$^{7}$, and Tom~Theuns$^{6}$ \\
$^{1}$Department of Astronomy and Theoretical Astrophysics Center, University of California, Berkeley, CA 94720-3411, USA \\
$^{2}$Academia Sinica Institute of Astronomy and Astrophysics, PO Box 23-141, Taipei 10617, Taiwan \\
$^{3}$Heidelberg Institute for Theoretical Studies, Schloss-Wolfsbrunnenweg 35, 69118, Heidelberg, Germany \\
$^{4}$Astronomy Department, Yale University, PO Box 208101, New Haven, CT 06520-8101, USA \\
$^{5}$Max-Planck-Institut f\"ur Astrophysik, Karl-Schwarzschild Stra{\ss}e 1, PO Box 1317, 85748, Garching, Germany \\
$^{6}$Institute for Computational Cosmology, Physics Department, University of Durham, South Road, Durham DH1 3LE, UK \\
$^{7}$Leiden Observatory, Leiden University, PO Box 9513, 2300 RA, Leiden, the Netherlands \\
$^{8}$Astrophysics Research Institute, Liverpool John Moores University, 146 Brownlow Hill, Liverpool, L3 5RF, UK
}

\begin{document}

\date{Accepted 2016 December 21. Received 2016 December 13; in original form 2016 November 11}

\pagerange{\pageref{firstpage}--\pageref{lastpage}} \pubyear{2017}

\maketitle

\label{firstpage}

\begin{abstract}

Galaxies that have fallen into massive haloes may no longer be able to accrete gas from their surroundings, a process referred to as `starvation' or `strangulation' of satellites. We study the environmental dependence of gas accretion onto galaxies using the cosmological, hydrodynamical EAGLE simulation. We quantify the dependence of gas accretion on stellar mass, redshift, and environment, using halo mass and galaxy overdensity as environmental indicators. We find a strong suppression, by many orders of magnitude, of the gas accretion rate in dense environments, primarily for satellite galaxies. This suppression becomes stronger at lower redshift. However, the scatter in accretion rates is very large for satellites. This is (at least in part) due to the variation in halocentric radius, since gas accretion is more suppressed at smaller radii. Central galaxies are influenced less strongly by their environment and exhibit less scatter in their gas accretion rates. The star formation rates of both centrals and satellites show similar behaviour to their gas accretion rates. The relatively small differences between gas accretion and star formation rates demonstrate that galaxies generally exhaust their gas reservoir somewhat faster at higher stellar mass, lower redshift, and in denser environments. We conclude that the environmental suppression of gas accretion could directly result in the quenching of star formation. 

\end{abstract}


\section{Introduction}

Star-forming galaxies are observed to lie on a fairly tight relation between their star formation rate (SFR) and stellar mass and there appears to be an equilibrium between gas supply and gas removal \citep[e.g.][]{Noeske2007, Bouche2010, Schaye2010, Lee2015}. Quenched galaxies are gas-poor and have SFRs significantly below this `star-forming main sequence'. The mechanisms proposed to be responsible for such low gas masses and SFRs include the removal of the galaxy's interstellar medium (ISM) as well as the prevention of further gas accretion onto the galaxy. 

The ISM can be removed through feedback from star formation or supermassive black holes or through environmental effects, such as ram pressure, tidal stripping, or satellite-satellite encounters. When the ISM is removed efficiently, this results in a fast decline of the SFR. The removal of a galaxy's gaseous halo through ram pressure can result in the suppression of gas accretion onto the galaxy and a slow decline of the SFR, also known as `starvation' or `strangulation' \citep[e.g.][]{Larson1980, Balogh2000}. All of these mechanisms are thought to play a role in the quenching of galaxies, but which one dominates likely depends on stellar mass and environment \citep[e.g.][]{Bosch2008, Fabello2012, Bahe2015}. In order to keep galaxies quenched for an extended period, resulting in the observed red colours, their gas accretion needs to be halted or significantly reduced compared to that of blue, star-forming galaxies. This may be caused by a different process than the one that quenches galaxies in the first place or there could be multiple quenching mechanisms acting together \citep[e.g.][]{Bahe2015, Marasco2016, Trayford2016}.

Observationally, galaxy properties such as colour, morphology, and specific SFR have been shown to depend on stellar mass \citep[e.g.][]{Kauffmann2003, Baldry2004, Peng2010, Moustakas2013}. Similarly, there is a clear dependence of these galaxy properties on environment, at least at low redshift \citep[e.g.][]{Dressler1980, Kauffmann2004, Peng2010, Wetzel2012}. Whether or not similar environmental dependencies exist at $z>1$, where the cosmic SFR density is significantly higher, is still an open question, but most studies find weaker environmental trends \citep[e.g.][]{Scoville2013, Lin2016}. 

The environment of a galaxy is thought to be primarily determined by properties of the dark matter dominated halo it lives in \citep[e.g.][]{Crain2009} and commonly used environmental indicators all correlate strongly with halo mass \citep{Haas2012, Marasco2016}. Indeed, using observed galaxy group catalogues, it has been shown that colour and star formation history correlate mostly with the mass of the host halo \citep[e.g.][]{Blanton2007, Wilman2010}. The virial radius, $R_\mathrm{vir}$, of a halo approximately marks the location of an accretion shock, within which the shocked gas is in virial equilibrium with the dark matter. For sufficiently massive systems, the gas temperature in the halo is much higher than that in the unshocked intergalactic medium. The temperature can be further increased through galactic feedback \citep[e.g.][]{VoortSchaye2012, Voortetal2016, Fielding2016}. Central galaxies sit at the minimum of the host halo's potential well, where the gas densities and cooling rates are highest, whereas satellite galaxies have fallen into more massive haloes and are thus offset from the centre of the potential.

The quenching mechanisms ram pressure stripping and starvation are special in the sense that they only operate on satellite galaxies. When a galaxy falls into a massive halo, it can lose its own gaseous halo and thus lose the reservoir from which it can replenish its ISM \citep[e.g.][]{Larson1980, Balogh2000, Peng2015}. Starvation therefore results in relatively slow quenching as the galaxy slowly consumes its cold gas reservoir through star formation. When the pressure of the external hot gas is high enough to also remove the ISM of the satellite galaxy, this is referred to as ram pressure stripping \citep[e.g.][]{Gunn1972, Quilis2000, Roediger2007, Tonnesen2009, Steinhauser2016} and results in relatively fast quenching of star formation. 

Gas accretion, or the lack thereof, is difficult to observe directly, but it affects the SFR, which is a key observable for galaxies. Additionally, a decrease of the infall of pristine gas can explain the fact that the metallicity is observed to increase with decreasing SFR at fixed stellar mass \citep{Mannucci2010, Dave2011, Peng2015, Lagos2016, Kacprzak2016, Bahe2017}. To directly study gas accretion, cosmological, hydrodynamical simulations are a powerful tool and they enable us to quantify the effect of starvation in different environments. \citet{Simha2009} indeed found lower gas accretion rates for satellites in massive haloes compared to centrals of the same mass. The current generation of state-of-the-art cosmological simulations has both the resolution and the range of environments required to study the dependence of gas accretion on halo mass, stellar mass, and galaxy overdensity in detail. This provides an important means by which to understand the quenching of galaxies in general and of satellites in particular.

In this paper, we use the EAGLE simulation \citep{Schaye2015, Crain2015} to study the gas accretion rates of central and satellite galaxies, and how they depend on stellar mass, environment, and redshift. Relatedly, the removal of cold gas through ram pressure and tidal stripping in the same simulation was studied by \citet{Marasco2016}, who showed that EAGLE reproduces the observed H\,\textsc{i} mass--environment trends. 

In Section~\ref{sec:sim} we describe the simulation used and the way we define galaxies and haloes (Section~\ref{sec:gal}) and gas accretion onto galaxies (Section~\ref{sec:acc}). In Section~\ref{sec:results} we present our results. We start by using halo mass as the environmental indicator, showing results for central and satellite galaxies separately in Section~\ref{sec:mass}. Section~\ref{sec:rad} shows how accretion rates vary with halocentric distance. We then adopt a geometric definition for environment (as often used in observations) in Section~\ref{sec:env}, namely galaxy overdensity, based on the 10th nearest neighbour, which confirms our previous results, albeit with small modifications. In Section~\ref{sec:sfr} we show the implications for the resulting SFRs. The evolution with redshift is discussed throughout Section~\ref{sec:results}. Finally, we discuss and summarize our results in Section~\ref{sec:concl}.

\section{Method} \label{sec:sim}

This work is part of the Evolution and Assembly of GaLaxies and their Environments (EAGLE) project \citep{Schaye2015, Crain2015}, which consists of a large number of cosmological simulations with varying subgrid physics, size, and resolution. The simulations use an extensively modified version of \textsc{gadget-3} \citep[last described in][]{Springel2005}, a smoothed particle hydrodynamics (SPH) code. The main modifications are the formulation of SPH, the time stepping and the subgrid physics. For our main results we make use of the largest-volume simulation (\emph{``Ref-L100N1504''}). The model is comprehensively described in \citet{Schaye2015} and we will summarize its main properties here.

The simulations assume a $\Lambda$CDM cosmology with parameters $\Omega_\mathrm{m} = 1 - \Omega_\Lambda = 0.307$, $\Omega_\mathrm{b} = 0.04825$, $h = 0.6777$, $\sigma_8 = 0.8288$, $n = 0.9611$ \citep{Planck2014}. A cubic volume with periodic boundary conditions is defined, within which the mass is distributed over $1504^3$ dark matter and an equal number of gas particles. The volume of the simulation used in this work is (100~comoving~Mpc)$^3$. The (initial) particle masses for baryons and dark matter are $1.8\times10^6$~M$_\odot$ and $9.7\times10^6$~M$_\odot$, respectively. The gravitational softening length is 2.66~comoving~kpc, but is limited to a maximum value of 0.7~proper~kpc, which is reached at $z=2.8$.

For gas with hydrogen number densities $n_\mathrm{H}\ge10^{-5}$~cm$^{−3}$ a temperature floor of 8000~K is imposed. Additionally, a temperature floor $T_\mathrm{EoS}$ is imposed at all densities. This takes the form of a polytropic equation of state $P_\mathrm{tot}\propto\rho_\mathrm{gas}^{4/3}$, where $P_\mathrm{tot}$ is the total pressure and $\rho_\mathrm{gas}$ is the density of the gas, normalized to $T_\mathrm{EoS}=8000$~K at $n_\mathrm{H}=0.1$~cm$^{-3}$. 

In dense gas, star formation is modelled according to the recipe of \citet{Schaye2008}. The SFR per unit mass depends on the gas pressure and reproduces the observed Kennicutt-Schmidt law \citep{Kennicutt1998} by construction. Following \citet{Vecchia2012}, gas is eligible to form stars when it has a temperature within 0.5~dex of $T_\mathrm{eos}$ and density exceeding $n_\mathrm{H}^\star=0.1$~cm$^{-3} \Bigl(\dfrac{Z}{0.002}\Bigr)^{-0.64}$, where $Z$ is the gas metallicity. Because the true, multiphase structure of the ISM is not resolved, the gas temperature reflects the effective pressure and the density should be interpreted as the mean density of the ISM. In this work, the galaxies' ISM is therefore defined to be all of the star-forming gas (i.e.\ all particles with $\mathrm{SFR}>0$).

The abundances of eleven elements released by massive stars and intermediate mass stars are followed as described in \citet{Wiersma2009b}. We assume the stellar initial mass function (IMF) of \citet{Chabrier2003}, ranging from 0.1 to 100~M$_\odot$. Radiative cooling and heating are computed element by element in the presence of the cosmic microwave background radiation and the \citet{Haardt2001} model for the UV/X-ray background from galaxies and quasars, assuming the gas to be optically thin and in (photo-)ionization equilibrium, as in \citet{Wiersma2009a}.

Feedback from star formation is implemented using the prescription of \citet{Vecchia2012} by injecting energy in thermal form. Neighbouring gas particles are selected stochastically (depending on the available energy) and then heated by a fixed temperature increment of $10^{7.5}$~K. The amount of energy available depends on the local gas density and metallicity and its parameters were chosen to approximately reproduce the observed $z=0$ galaxy stellar mass function and galaxy sizes \citep{Schaye2015, Crain2015}.

The model for the formation of and feedback from supermassive black holes is fully described in \citet{Rosas2015} and \citet{Schaye2015}. A seed mass black hole of about $1.4\times10^5$~M$_\odot$ is placed in every halo with mass $M_\mathrm{halo}\gtrsim1.4\times10^{10}$~M$_\odot$ that does not already contain a black hole \citep{Springeletal2005}. These black holes grow by the accretion of gas, after which energy is injected into the surrounding medium, and by mergers. A fraction of 1.5 per cent of the rest-mass energy of the gas accreted onto the black hole is injected into the surrounding medium in the form of heat, by increasing the temperature of at least one neighbouring gas particle by at least $10^{8.5}$~K. Parameter values were chosen to obtain a good match with observed galaxy masses and sizes as well as with the $z=0$ relation between galaxy mass and black hole mass.\citep{Crain2015}.

These simulations were not tuned to reproduce any observable gas properties. However, they have been shown to reproduce a number of observations reasonably well, such as the H\,\textsc{i} column density distribution and covering fraction \citep{Rahmati2015}, the H\,\textsc{i} and H$_2$ content of galaxies with stellar mass above $10^{10}$~M$_\odot$ \citep{Lagos2015, Bahe2016, Crain2017}, and the dependence of the H\,\textsc{i} mass on environment \citep{Marasco2016}.
However, the predicted soft X-ray luminosity and halo gas mass fraction in groups and clusters are somewhat higher than observed \citep{Schaye2015}, whilst there may not be sufficient cool gas in low-mass galaxies \citep{Crain2017}.
The simulation also reproduces the observed colour-magnitude diagram \citep{Trayford2015}, the FIR properties of galaxies \citep{Camps2016}, their optical colours (Trayford et al. submitted), alpha enhancements \citep{Segers2016b}, and the evolution of the galaxy mass function \citep{Furlong2015} and galaxy sizes \citep{Furlong2017}.

\subsection{Identifying haloes and galaxies} \label{sec:gal}

In order to identify haloes, we first find dark matter haloes using a Friends-of-Friends (FoF) algorithm. If the separation between two dark matter particles is less than 20 per cent of the average dark matter particle separation, they are placed in the same group. Baryonic particles are placed in a group if their nearest dark matter neighbour is part of the group. We then apply \textsc{subfind} \citep{Springel2001, Dolag2009} to the FoF output to find the gravitationally bound particles and to identify subhaloes. The galaxy containing the minimum of the gravitational potential is called the central. Note that when the central and satellite are of very similar mass, there is some ambiguity in the assignment of either label. 

In this paper, we use the most bound particle of a halo or subhalo, as found by \textsc{subfind}, as the halo or galaxy centre. We then use a spherical overdensity criterion, considering all the particles in the simulation, to determine the halo mass, $M_\mathrm{halo}$, of every host halo (thus excluding subhaloes). Here we define $M_\mathrm{halo}$ to be the total mass enclosed by a radius, $R_\mathrm{vir}$, within which the mean overdensity is 200 times the mean density of the Universe at its redshift. Satellites are subhaloes of a larger structure identified by \textsc{subfind} and may or may not have their own dark matter halo. We include only those that reside within $R_\mathrm{vir}$ of their host halo in order to exclude spurious satellite identifications, but our results are insensitive to this choice. Stellar mass, $M_\mathrm{star}$, is measured using the star particles associated with the galaxy's subhalo, but only including those within 30~proper kpc of the galaxy's centre in order to exclude intrahalo stars.  For each galaxy, we calculate the ISM mass by adding up the mass of star-forming gas particles in the galaxy's subhalo, again within 30~kpc of the centre.

\subsection{Calculating gas accretion onto galaxies} \label{sec:acc}

\begin{figure*}
\center
\includegraphics[scale=.75]{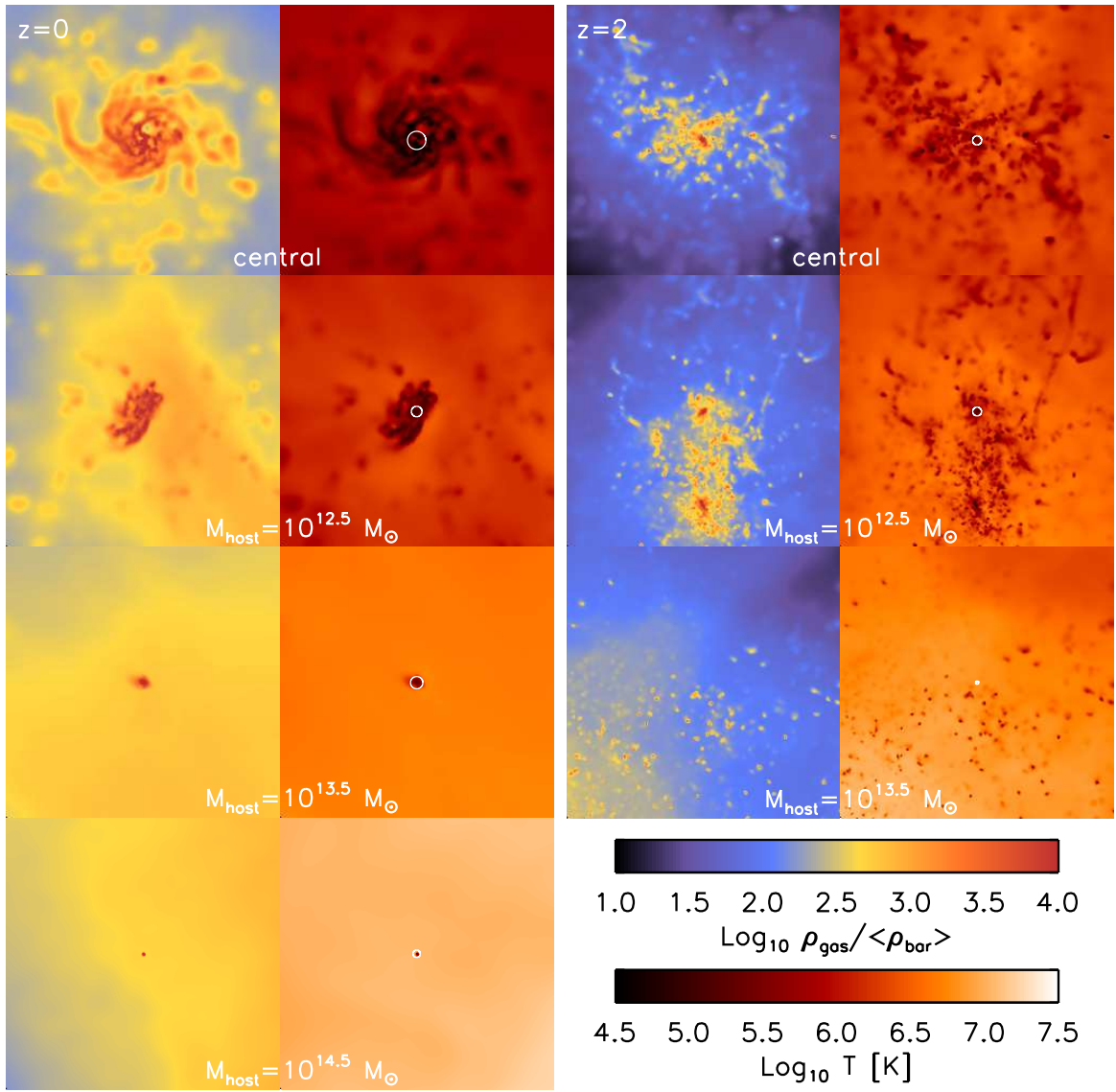}
\caption {\label{fig:img} Projected (200~proper~kpc)$^3$ images of the average gas overdensity (first and third columns) and temperature (second and fourth columns) at $z=0$ (left) and $z=2$ (right), each centred on a randomly chosen galaxy with $M_\mathrm{star}=10^{10}$~M$_\odot$. White circles in the temperature panels indicate the galaxies' stellar half-mass radii. The top row shows central galaxies, while the other rows show satellite galaxies for which the host halo mass increases from $10^{12.5}$~M$_\odot$ (second row) to $10^{13.5}$~M$_\odot$ (third row) to $10^{14.5}$~M$_\odot$ (bottom row). The temperature of the surrounding gas increases with host halo mass, which increases the effect of the environment on gas accretion rates. At high redshift, there is much more cold, dense gas around the galaxies. Gas at the same overdensity also has a factor of 27 higher density at $z=2$ than at $z=0$, because of the cosmological expansion of the Universe. Together these effects result in a decrease of the environmental dependence with increasing redshift.}
\end{figure*}

Gas accretion at $z_1>z>z_2$ is calculated in the following way. We select the gas particles that were not part of any galaxy's ISM at $z_1$, i.e.\ those that were not star-forming. This automatically excludes accretion of ISM gas through galaxy mergers, but still allows accretion of non-star-forming gas previously surrounding external galaxies. Of those selected particles, we consider the gas and recently formed star particles that are in the galaxy of interest at $z_2$ (and were non-star-forming gas at $z_1$) as having been accreted between $z_1$ and $z_2$. Gas particles that accrete after $z_1$, but are ejected or stripped before $z_2$ are therefore not included in our accretion rates. This affects only the normalization of our results, but not the trends. When approximately doubling the time interval between $z_1$ and $z_2$, we find that the normalization of the gas accretion rates changes by 0.1~dex or less. We therefore conclude that the contribution from accreted-but-rapidly-ejected gas is likely subdominant.
We include all accreted gas, regardless of its history, which therefore includes accretion from an outer non-star-forming gas disc as well as direct accretion from the gaseous halo.  
The gas accretion rate differs from the change in gas mass even in the absence of star formation and stellar mass loss, because the removal of gas that is in the ISM at $z_1$, through stripping or outflows, is not taken into account. 

For brevity, accretion at $0.1>z>0$ ($2.24>z>2$) will be referred to as accretion at $z=0$ ($z=2$). This redshift interval corresponds to 1.35~Gyr (0.33~Gyr) in our assumed cosmology. The minimum non-zero accretion rate we can measure for an individual galaxy is therefore $1.3\times10^6$~M$_\odot$~Gyr$^{-1}$ at $z=0$ and $5.5\times10^6$~M$_\odot$~Gyr$^{-1}$ at $z=2$ (i.e.\ one accreted gas particle over the time interval). Halo mass or host mass refers to the mass of the host halo in which the satellite resides at $z_2$, unless otherwise stated. We repeated our analysis at $1.26>z>1$ and found it to show behaviour intermediate between $z=0$ and $z=2$, although more similar to $z=2$, and therefore chose not to include it in this paper.

\section{Results} \label{sec:results}


As described in the introduction, gas accretion may proceed very differently for satellite and central galaxies. Satellites may be stripped of their own haloes and no longer able to replenish their ISM through accretion. In this section we investigate the importance of this starvation and its dependence on stellar mass and redshift, as well as on environment, which we separate into a dependence on halo mass (Section~\ref{sec:mass} and~\ref{sec:sfr}), halocentric radius (Section~\ref{sec:rad}), and galaxy overdensity (Section~\ref{sec:env}).

Regarding gas accretion onto galaxies, an important difference between high and low redshift is the density of the circumgalactic medium, which controls the cooling rate, and thus affects the amount of cold, dense gas. At fixed redshift, the gas accretion rate is affected by the environment through differences in the temperature and pressure of the hot, diffuse gas and the amount of cold, dense gas around a galaxy. Figure~\ref{fig:img} shows these gas properties visually. Each image is a projection of the same physical volume, (200~proper~kpc)$^3$. The galaxy in the centre of each image has approximately the same stellar mass, $M_\mathrm{star}=10^{10}$~M$_\odot$, and its stellar half-mass radius (computed within 30~proper kpc) is indicated by a white circle. The first and third columns show volume-weighted average gas overdensity and the second and fourth columns show mass-weighted average gas temperature, at $z=0$ (left) and $z=2$ (right). The images in the top row show central galaxies and the other three rows show satellite galaxies for which the host halo mass increases from $M_\mathrm{host}=10^{12.5}$~M$_\odot$ (second row) to $10^{13.5}$~M$_\odot$ (third row) to $10^{14.5}$~M$_\odot$ (bottom row, only at $z=0$). The satellites' (proper) halocentric distances are 186~kpc ($10^{12.5}$~M$_\odot$), 318~kpc ($10^{13.5}$~M$_\odot$), and 710~kpc ($10^{14.5}$~M$_\odot$) at $z=0$ and 108~kpc ($10^{12.5}$~M$_\odot$) and 205~kpc ($10^{13.5}$~M$_\odot$) at $z=2$.

We selected these galaxies randomly, but verified that others show very similar gas properties (except that the mean density and, to a lesser extent, the mean temperature depend on the distance from the centre of the halo). The gaseous environments of central galaxies and satellites in $10^{12.5}$~M$_\odot$ host haloes are similar (first and second row in Figure~\ref{fig:img}), so $M_\mathrm{star}=10^{10}$~M$_\odot$ satellites may not be affected much by their environment. However, the temperature of the surrounding gas increases strongly with host halo mass (third and fourth row), which also means that the average pressure and the average cooling time increase (at fixed density). This makes it harder for a satellite to maintain its own gaseous halo and to accrete cold gas onto its ISM. 

As expected, there is a large difference between high and low redshift, with a lot more cold, dense gas surrounding the galaxies at $z=2$ in all environments. These denser satellite haloes will be less easily stripped. Note that there is a factor of 27 difference in physical density between gas at fixed overdensity at $z=2$ and $z=0$, because the average density of the universe decreases with decreasing redshift. At higher redshift, a larger fraction of the accreting gas never shock heats to the virial temperature near the virial radius of the halo, known as `cold-mode' accretion (e.g.\ \citealt{Keres2005, Dekel2009a, Voort2011a}, Correa et al.\ in preparation). Average (gas and dark matter) densities are higher and cooling times are shorter, which result in higher gas accretion rates onto a galaxy. The availability of gas with short cooling times likely reduces the effect of the (hot halo) environment.

\subsection{Dependence on halo mass} \label{sec:mass}

\begin{figure*}
\center
\includegraphics[scale=.52]{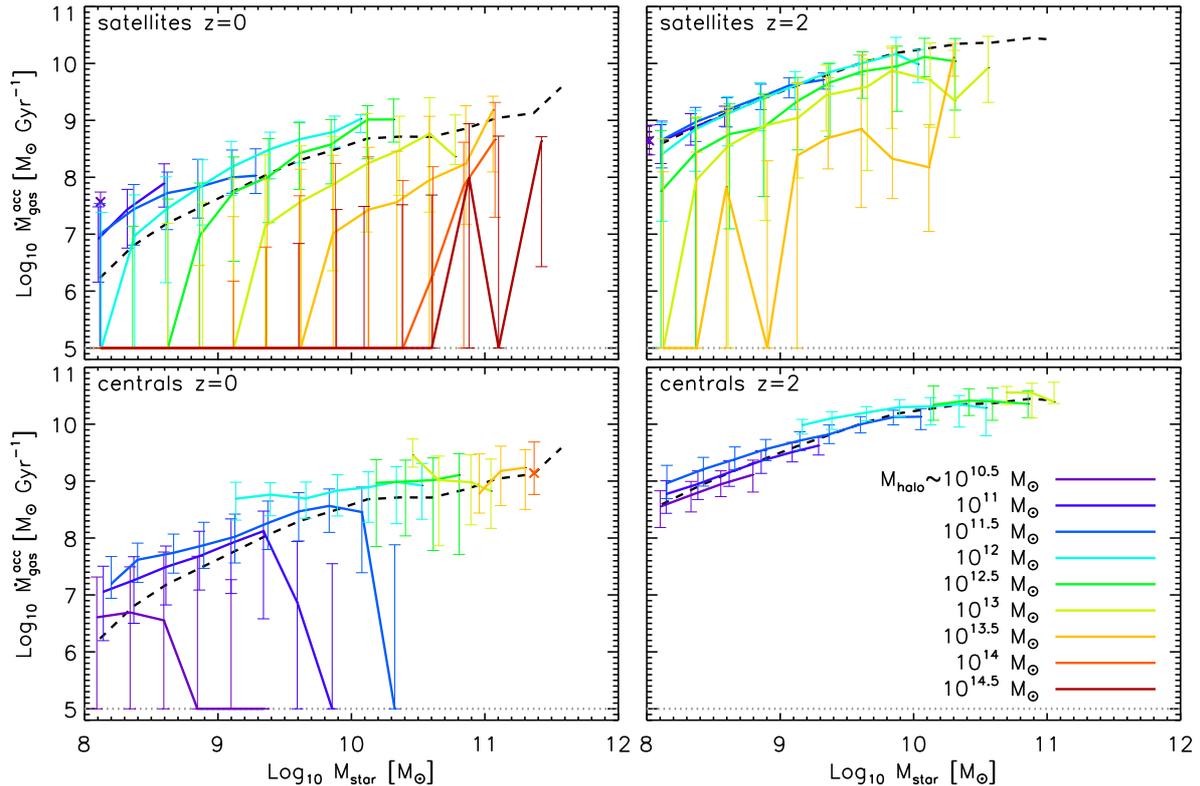}
\caption {\label{fig:mass} Median gas accretion rate onto galaxies as a function of stellar mass for satellite (top) and central galaxies (bottom) at $z=0$ (left) and $z=2$ (right). The dashed curves show median accretion rates for all galaxies. Solid curves and crosses show median accretion rates in halo mass bins, 0.5~dex in width, centred on the values indicated in the legend. Gas accretion rates of zero have been set to $10^5$~M$_\odot$~Gyr$^{-1}$, as shown by the dotted, grey line. Error bars show the 16th and 84th percentiles of the distribution. Only bins containing at least 5 galaxies are shown. Gas accretion onto satellites is more strongly suppressed in more massive host haloes and for lower-mass galaxies. Although much weaker, this trend is reversed for central galaxies. The environmental dependence becomes stronger and the scatter becomes larger at lower redshift.}
\end{figure*}

The environmental effects on gas accretion, discussed qualitatively in the previous section, are shown quantitatively in Figure~\ref{fig:mass}. The median gas accretion rate onto galaxies is shown as a function of stellar mass for satellite (top) and central galaxies (bottom) at $z=0$ (left) and $z=2$ (right). The dashed curves show median accretion rates for all galaxies and are thus identical in the top and bottom panels. Solid curves show median accretion rates in halo mass bins, 0.5~dex in width, centred on the values indicated in the legend and increasing from purple ($10^{10.5}$~M$_\odot$) to red ($10^{14.5}$~M$_\odot$). This halo mass is the mass of the host halo in which the satellite resides; it is not the mass of the satellite's subhalo. For visualization purposes, gas accretion rates of zero have been set to $10^5$~M$_\odot$~Gyr$^{-1}$ (as also indicated by the grey, dotted line). Note that we cannot detect accretion of less than 1 particle per redshift interval, so the `zero' rates could in reality be higher ($<1.3\times10^6$~M$_\odot$~Gyr$^{-1}$ at $z=0$ and $<5.5\times10^6$~M$_\odot$~Gyr$^{-1}$ at $z=2$). Thin, coloured error bars show the $1\sigma$ scatter (i.e.\ 16th and 84th percentiles of the distribution). Only bins containing at least 5 galaxies are shown. 

In general, the gas accretion rate increases with stellar mass, a consequence of the increase of the dark matter and gas accretion rate onto haloes \citep[e.g.][]{Neistein2006, Fakhouri2010, Voort2011a, Correa2015}. It increases more strongly with stellar mass at the low-mass end than at the high-mass end. This can be explained by the increase of the cooling time of the circumgalactic medium at higher masses (with higher virial temperatures) and by the mass dependence of stellar and active galactic nucleus (AGN) feedback \citep[e.g.][]{Voort2011a}, which sets the shape of the galaxy stellar mass function. 

Satellites of fixed stellar mass accrete gas at rates that differ by many orders of magnitude, depending on the mass of their host halo. At fixed stellar mass, the suppression of gas accretion becomes gradually stronger with increasing halo mass. This is due to the fact that the temperature and pressure of the halo gas as well as the velocity of the satellite increase with mass, which makes the removal of the satellite's gaseous subhalo more efficient and the accretion of gas onto the satellite more difficult. Conversely, at fixed halo mass, the suppression of gas accretion becomes weaker with increasing stellar mass. This is due to the fact that the temperature and pressure difference between the satellite's subhalo and its host halo are smaller and that the potential well of the satellite is deeper, which makes stripping of the satellite's subhalo less efficient and gas accretion easier. Lower-mass satellites have also been shown to spend more time in their host halo before they merge or disrupt due to the reduced dynamical friction. This means that low-mass satellites at a given redshift have on average spent more time as satellites, increasing the likelihood of their subhaloes being stripped away \citep[e.g.][]{Boylan2008, Wetzel2010}. Similar trends with stellar mass and environment are seen for the H\,\textsc{i} content of satellites \citep{Marasco2016}.

On the other hand, central galaxies behave very differently. The dependence on halo mass at fixed stellar mass is much weaker and the gas accretion rate is slightly higher in more massive haloes (opposite to the trend for satellites). Since the central resides at the minimum of the potential, gas cooling and accretion can happen efficiently and increase with halo mass. The higher gas accretion rate will lead to a higher SFR and thus, eventually, to a higher stellar mass. In this way, central galaxies regulate their own growth, which keeps the scatter in the `star-forming main sequence' and stellar mass--halo mass relation small. In general, the accretion rate increases with stellar mass, but the median accretion rate is very low for the more massive galaxies in the three lowest halo mass bins ($M_\mathrm{halo}\lesssim10^{11.5}$~M$_\odot$). These haloes' gas reservoirs are too small to efficiently feed the central galaxy. Additionally, these galaxies have formed more stars than lower-mass galaxies in similar haloes and thus produced stronger outflows, which prevents accretion at least temporarily. 

The scatter in gas accretion rates is very large for galaxies that are strongly influenced by their environment. This results (at least in part) from the fact that some satellites will have fallen into the host halo recently and have not yet been affected, whereas others have spent considerably more time inside the halo and were stripped of their own halo (and possibly their ISM). This time dependence will be studied by Bah\'e et al.\ (in preparation). This means, however, that just knowing the stellar mass and host halo mass is not enough to predict a galaxy's accretion rate. Observationally, galaxies in cluster environments show very large range of SFRs at fixed stellar mass \citep{Chies2015, Paccagnella2016}, which potentially results from the large range in gas accretion rates for satellites.

As expected, gas accretion rates onto galaxies are higher at $z=2$ than at $z=0$, following increased dark matter and gas accretion rates onto haloes \citep[e.g.][]{Neistein2006, Fakhouri2010, Voort2011a, Correa2015}. The dependence on halo mass is weaker at $z=2$, because gas densities in the circumgalactic medium are higher at high redshift and the accretion of cold, dense gas is less readily affected by the environment than that of hot, diffuse gas. Additionally, dynamical friction is stronger, which means that galaxies merge more quickly at higher redshift and thus spend less time as satellites. Nonetheless, even at $z=2$, gas accretion onto satellites is substantially suppressed at $M_\mathrm{halo}\gtrsim10^{12.5}$~M$_\odot$. The trends with environment are similar to those at $z=0$, except for the absence of massive centrals in low-mass host haloes with low accretion rates. Observations show a reduced quenching of satellite galaxies towards higher redshift, which also indicates a smaller environmental effect \citep[e.g.][]{Kawinwanichakij2016, Nantais2016, Hatfield2016}.

\subsection{Radial dependence} \label{sec:rad}

\begin{figure}
\center
\includegraphics[scale=.52]{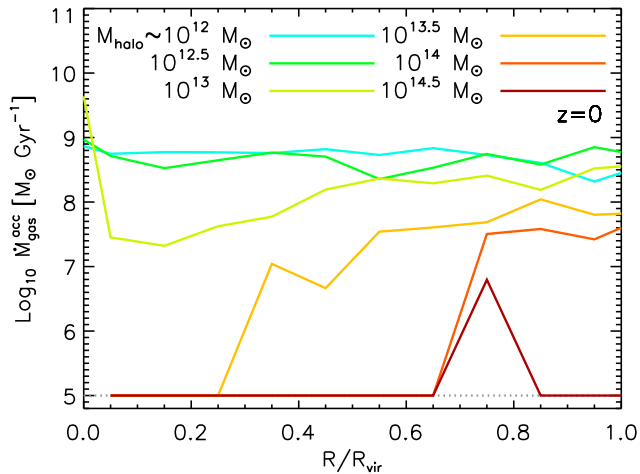}
\caption {\label{fig:rad} Median $z=0$ gas accretion rate onto galaxies with $10^{9.5}<M_\mathrm{star}<10^{10.5}$~M$_\odot$ as a function of 3D distance from the host halo centre, normalized by their host halo's virial radius. Different coloured curves correspond to different halo mass bins (0.5~dex wide) as indicated in the legend. Gas accretion rates of zero have been set to $10^5$~M$_\odot$~Gyr$^{-1}$, as shown by the dotted, grey line. Bins contain at least 5 galaxies and only central galaxies are included in the first radial bin. For high-mass haloes, the gas accretion onto galaxies decreases by orders of magnitude towards the centre of the host halo. This may indicate that environmental effects are stronger at smaller radii or that galaxies near the centre have spent more time in their host halo or both.}
\end{figure}

It is expected that the suppression of gas accretion due to the host halo environment, and thus the quenching of star formation, is a slow process \citep[e.g.][]{McCarthy2008, Simha2009, Wetzel2013}. On average, satellites in the outskirts of the halo have likely fallen in more recently than those in the centre of the host halo \citep{Gao2004}. This should then result in a radial dependence of the gas accretion rates with respect to the halo centre. Figure~\ref{fig:rad} shows the $z=0$ median gas accretion rate for galaxies with $10^{9.5}<M_\mathrm{star}<10^{10.5}$~M$_\odot$ as a function of 3D distance from the host halo centre, normalized by the virial radius of the halo. The environmental dependence is shown by different curves, which include haloes within 0.25~dex of the values indicated in the legend.  For clarity, median gas accretion rates of zero are shown at $10^5$~M$_\odot$~Gyr$^{-1}$. Note that this calculation uses 3D radii while any observed radial dependence would be diluted by projection effects. The radial trends are similar for galaxies with lower (higher) stellar masses than $\approx10^{10}$~M$_\odot$, but the suppression of the gas accretion rate at fixed halo mass and radius is stronger (weaker), as expected from Figure~\ref{fig:mass}. Similar, but weaker, radial trends also exist at higher redshift. 

The two lowest halo mass bins that contain $M_\mathrm{star}\approx10^{10}$~M$_\odot$ satellite galaxies (cyan and green curves; $M_\mathrm{halo}\approx10^{12-12.5}$~M$_\odot$) show no radial dependence of the gas accretion rates onto satellite galaxies. This is consistent  with Figure~\ref{fig:mass}, where we found little suppression of gas accretion for these masses. However, a strong radial dependence can be seen for $M_\mathrm{halo}\gtrsim10^{13}$~M$_\odot$, with the median accretion rate decreasing towards smaller radii. At fixed (normalized as well as physical) radius, the accretion rate is lower in more massive haloes and the radius where the median accretion rate reaches zero in our simulations moves to higher values. This radial dependence may indicate that environmental effects are stronger at smaller halocentric radii or that galaxies near the centre have spent more time in their host halo or both. Knowing the position with respect to the centre of the halo can thus help reduce the scatter in gas accretion rate at fixed stellar and halo mass. 

In our most massive haloes ($M_\mathrm{halo}\gtrsim10^{13.5}$~M$_\odot$), the accretion rate of these satellites is, on average, suppressed throughout the entire halo and it is possible that the halo's influence extends beyond its virial radius. The dependence outside $R_\mathrm{vir}$ will be studied in Bah\'e et al.\ in preparation. 

If starvation is a major contributor to satellite quenching, this radial dependence of the gas accretion rate will result in a similar radial dependence of the SFR. \citet{Paccagnella2016} indeed find that, at fixed stellar mass, the SFRs are lower in cluster centres than in their outskirts (see also \citealt{Wetzel2012}). Although this is consistent with our results, other environmental quenching mechanisms, such as ram pressure stripping or tidal stripping, would lead to similar behaviour.

\subsection{Dependence on galaxy overdensity} \label{sec:env}

\begin{figure*}
\center
\includegraphics[scale=.52]{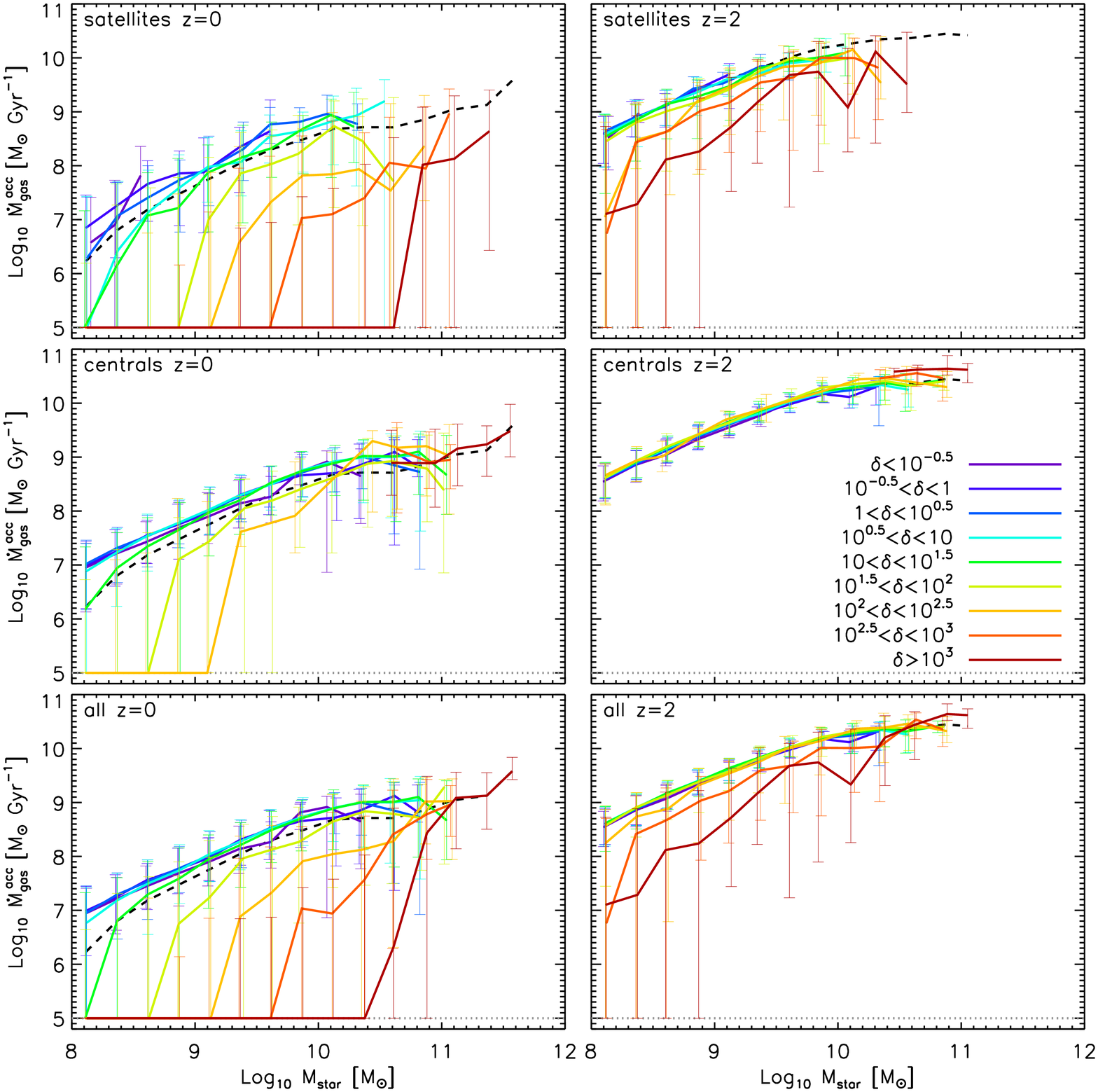}
\caption {\label{fig:env} Median gas accretion rates as a function of stellar mass for satellites (top), centrals (middle) and all galaxies (bottom) at $z=0$ (left) and $z=2$ (right). Dotted and dashed curves are the same as in Figure~\ref{fig:mass}. Solid curves show median gas accretion rates in bins of galaxy overdensity, $\delta\equiv\rho_{10}/\rho_\mathrm{all}$, as indicated in the legend. Gas accretion rates of zero have been set to $10^5$~M$_\odot$~Gyr$^{-1}$. Error bars show the 16th and 84th percentiles of the distribution. Only bins containing at least 5 galaxies are shown. There is a very strong dependence on galaxy overdensity at $z=0$ and a weaker dependence at $z=2$. This dependence is primarily driven by satellite galaxies, although there is some residual dependence for centrals at $z=0$.}
\end{figure*}

Halo mass can be inferred in observations through various methods, such as galaxy clustering, weak lensing, satellite kinematics, and abundance matching. However, most of these methods are indirect and may introduce biases. To facilitate comparison to observations, we use the galaxy's 10th nearest neighbour to define environment in this section. This quantity is more directly available in galaxy surveys, although we measure the 3D galaxy density, whereas observations will necessarily use the projected 2D density and a relative velocity cut along the line-of-sight. Our resulting galaxy density, $\rho_{10}$, is defined as the number density of galaxies above our resolution limit (i.e.\ $M_\mathrm{star}>10^{8}$~M$_\odot$). The results do not change substantially if instead we choose to use e.g.\ the 5th nearest neighbour or $M_\mathrm{star}>10^{9}$~M$_\odot$. We then compute the average galaxy density in the full simulation volume, $\rho_\mathrm{all}$, and define the galaxy overdensity as $\delta\equiv\rho_{10}/\rho_\mathrm{all}$ and use this as our environmental parameter.

Figure~\ref{fig:env} is similar to Figure~\ref{fig:mass} but uses $\delta$ instead of $M_\mathrm{halo}$ to characterize a galaxy's environment. It shows median gas accretion rates as a function of stellar mass for satellites (top), centrals (middle) and all galaxies (bottom) at $z=0$ (left) and $z=2$ (right). In all panels, the dashed curves show median gas accretion rates for all galaxies in our simulations. Solid curves show median gas accretion rates in bins of galaxy overdensity, as indicated in the legend, and error bars show the $1\sigma$ scatter. 

The conclusions drawn from this definition of environment, i.e.\ galaxy overdensity, are very similar to those drawn above based on halo mass. This is expected, because galaxy overdensity correlates with halo mass \citep[e.g.][]{Haas2012, Marasco2016}. Considering all galaxies first (bottom panels), the gas accretion rate is lower in denser environments at fixed stellar mass. At fixed galaxy overdensity, the gas accretion rate is more suppressed for lower stellar masses. Note, however, that the scatter in accretion rate can be very large, likely due in part to the varying amounts of time a galaxy spent in a particular environment. The top panels show that this environmental effect is primarily driven by satellites, as also seen in observations of the environmental dependence of the quenched galaxy fraction \citep{Peng2012}. We find that the environmental effect decreases with increasing redshift (see also Section~\ref{sec:mass}). 

\begin{figure*}
\center
\includegraphics[scale=.52]{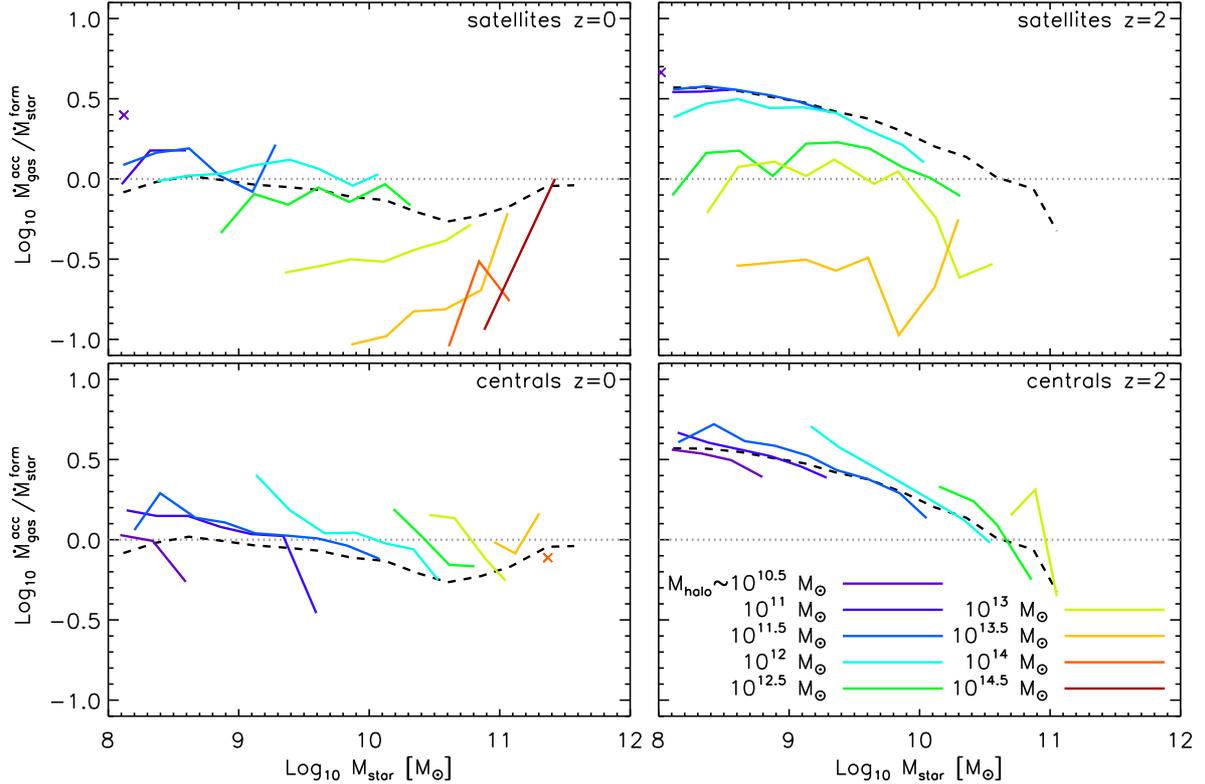}
\caption {\label{fig:sfr} The median gas accretion rate divided by the median SFR as a function of stellar mass for satellites (top) and central galaxies (bottom) at $z=0$ (left) and $z=2$ (right). The dashed curves show the gas-accretion-rate-to-SFR ratios for all galaxies. Solid curves and crosses show the gas-accretion-rate-to-SFR ratio in halo mass bins, 0.5~dex in width, centred on the values indicated in the legend. Only bins containing at least 5 galaxies and non-zero median gas accretion rates are shown. The dotted line shows a ratio of unity, where the median gas accretion rate and the median SFR are equal. There is only a relatively weak dependence of this ratio on stellar mass and environment, showing that the SFR is in large part driven by the gas accretion rate. In general, the ratio decreases with stellar mass, which means that massive galaxies exhaust their gas reservoirs more quickly than dwarf galaxies at $z=0$ and build up their gas reservoir more slowly at $z=2$. In denser environments (in this case in more massive host haloes), satellite galaxies have lower gas-accretion-rate-to-SFR ratios, by up to 1~dex, whereas centrals have slightly higher ratios.}
\end{figure*}

Satellites are affected by their environment more than centrals, but the effect on centrals is non-zero for large overdensities, $\delta\gtrsim30$, at $z=0$. We did not see this gas accretion suppression for centrals when we used halo mass as a probe of environment. Although the similarity with Figure~\ref{fig:mass} shows that halo mass is responsible for the main environmental effect in satellite galaxies, this implies a (smaller) residual environmental dependence besides halo mass. This may be due to the fact that, in dense environments, nearby galaxies are competing for the same gas reservoir and accrete gas that would otherwise have been available to other galaxies \citep{Aragon2014}. Relatedly, there may be environmental effects beyond $R_\mathrm{vir}$, either because the halo affects the intergalactic medium beyond $R_\mathrm{vir}$ or because a satellite's orbit brings it beyond $R_\mathrm{vir}$ rendering it a central once more (i.e.\ splashback galaxies). Additionally, galaxies in dense environments formed earlier, also known as assembly bias \citep[e.g.][]{Sheth2004, Gao2005}, which reduces their growth rate at later times. 

The environmental dependence of the gas accretion rate for central galaxies in our simulations is possibly reversed at the highest stellar masses at $z=2$, although this is a very small effect. Centrals at fixed stellar mass $>10^{10.5}$~M$_\odot$ have slightly higher gas accretion rates in higher density environments. This reversal of the trend could lead to a reversal in the SFR--density relation, which has been suggested observationally at $z=1-1.5$ \citep{Elbaz2007, Lin2016}, but only if the galaxies are predominantly centrals, which is likely the case for massive galaxies.

\subsection{Effect on star formation} \label{sec:sfr}


SFRs are primarily determined by the amount of gas in the ISM, which is replenished through gas accretion, gas-rich galaxy mergers, and stellar mass loss. It is difficult to measure gas accretion rates observationally, but SFRs are routinely derived. Because gas accretion and star formation are so intricately linked, we turn our attention to the balance between them and the dependence on environment.

The environmental dependence of star formation (and ISM mass) is very similar to the environmental dependence of gas accretion. At fixed environment (either halo mass or galaxy overdensity), the gas accretion rate tends to be somewhat more suppressed than the SFR. This is due to the fact that there is some time delay during which galaxies consume their existing ISM, before they are affected by starvation. Moreover, stellar mass loss can replenish the ISM even in the absence of accretion \citep{Leitner2011, Segers2016a}.

Instead of showing the environmental effect on star formation directly, which is qualitatively the same as the effect on gas accretion shown in previous figures, Figure~\ref{fig:sfr} shows the ratio of the median gas accretion rate and median SFR as a function of stellar mass for satellites (top) and centrals (bottom) at $z=0$ (left) and $z=2$ (right). The SFR is calculated as the stellar mass in a galaxy that was formed between $z=0.1$ ($z=2.24$) and $z=0$ ($z=2$) divided by the corresponding time interval. The dashed curves show gas-accretion-rate-to-SFR ratios for all galaxies. Solid curves show these ratios in halo mass bins, 0.5~dex in width, centred on the values indicated in the legend. We show only stellar mass bins where the median gas accretion rate is non-zero, because we cannot resolve low-level gas accretion and this would otherwise result in biased and unrealistic or undefined ratios. Although this excludes the lowest-stellar mass bins at fixed halo mass, no galaxies have been excluded for the stellar mass bins shown and our results are therefore not biased.  

The gas-accretion-rate-to-SFR ratio generally decreases with increasing stellar mass. At $z=2$, the most massive galaxies have ratios close to unity, on average, which implies that their ISM mass stays roughly constant in the absence of outflows, stripping, and mergers. Lower-mass galaxies, however, still have gas accretion rates higher than their SFRs and are building up their gas reservoirs. Low-mass galaxies at $z=0$ also have gas accretion rates that approximately balance their SFRs. Most $M_\mathrm{star}>10^{10}$~M$_\odot$ galaxies at $z=0$ are depleting their ISM. This is consistent with the downsizing trend that is found in observations and simulations, where more massive galaxies assemble their stellar mass earlier than less massive ones \citep[e.g.][]{Bundy2006, Perez2008, Crain2009, Muzzin2013, Okamoto2014, Voort2016}. At fixed halo mass, more massive centrals have lower gas-accretion-rate-to-SFR ratios. However, satellites in massive haloes ($M_\mathrm{halo}\gtrsim10^{13}$~M$_\odot$) at $z=0$ exhaust their ISM less quickly at higher stellar mass, because the effect of the environment on gas accretion and star formation decreases with increasing stellar mass. It is important to note that this comparison includes only gas accretion and star formation and excludes gas brought in by galaxy mergers and gas removed through outflows or stripping. The change in gas mass may therefore differ from the change implied by the gas-accretion-rate-to-SFR ratio.

The weaker dependence on stellar mass and halo mass (up to 1~dex) as compared to the gas accretion rate (Figure~\ref{fig:mass}) shows that the rate of gas accretion onto galaxies is one of the primary drivers of the SFR. Stripping of the ISM may also contribute or even dominate in certain regimes.
Even though gas accretion rates and SFRs show similar environmental dependences, the gas-accretion-rate-to-SFR ratio clearly shows a residual dependence on environment. 
At fixed stellar mass, satellites have lower gas-accretion-rate-to-SFR ratios if they reside in higher-mass haloes, indicating that their gas reservoir is building up more slowly ($M_\mathrm{halo}\lesssim10^{13}$~M$_\odot$ at $z=2$) or is being depleted more rapidly ($M_\mathrm{halo}\gtrsim10^{13}$~M$_\odot$ at $z=2$ and $M_\mathrm{halo}\gtrsim10^{12.5}$~M$_\odot$ at $z=0$). This is likely because of the time delay between the shut-down of gas accretion and that of star formation due to the remaining gas reservoir in the ISM and the replenishment by stellar mass loss. Centrals, on the other hand, exhibit a smaller environmental dependence of opposite sign. At fixed stellar mass, centrals in more massive haloes have slightly higher gas-accretion-rate-to-SFR ratios, which allows them to grow their ISM more rapidly, which (after some time delay) results in higher SFRs and higher stellar masses, keeping the scatter in the stellar mass--halo mass relation for central galaxies small \citep{Tinker2016, Matthee2016}.

\section{Discussion and conclusions} \label{sec:concl}

We have used the EAGLE simulation to study how the rate of gas accretion onto galaxies changes with environment. A galaxy's environment is defined either as the mass of the host halo in which it resides or as the local 3D galaxy overdensity based on its 10th nearest neighbour with $M_\mathrm{star}>10^8$~M$_\odot$. The most significant environmental effect is seen for satellites, whose gas accretion rates vary by many orders of magnitude (see Figures~\ref{fig:mass} and~\ref{fig:env}). At fixed stellar mass, the suppression of gas accretion is larger in more massive haloes and at higher galaxy overdensity, albeit with large scatter. This can be caused by the more extreme gas properties in higher-mass haloes, such as higher temperature and pressure and fewer cold, dense clumps, by the higher satellite velocities, and/or by the fact that satellites spend a larger amount of time in these harsher environments before they merge or disrupt. We also find a strong dependence on the halocentric distance for satellites, which have lower gas accretion rates at smaller radii (see Figure~\ref{fig:rad}). Similarly, this is likely also caused by the higher gas pressure in the halo centre, the higher velocities, and the longer time spent inside the halo. This radial dependence can also help explain the large scatter seen for satellites with similar stellar and host halo masses.

The effect of the environment decreases towards higher redshift, both at fixed halo mass and at fixed galaxy overdensity. Since the gas density is higher at higher redshift, the cooling time of the gas is shorter, and a larger fraction of the gas is cold and dense. Satellite galaxies are therefore less easily stripped of their own gaseous halo and can more easily accrete gas from their surroundings. Additionally, dynamical friction is stronger and galaxies thus merge more quickly at higher redshift, which causes them to spend less time as satellites, reducing environmental effects. 

Although satellites are affected more strongly by their environment, centrals are affected as well. At fixed stellar mass, centrals in more massive haloes have slightly higher gas accretion rates, opposite to the trend for satellites. Since more massive haloes have larger gas reservoirs, this results in higher accretion rates in their centres and also acts to keep the scatter in the halo-to-stellar mass relation low for central galaxies. Remarkably, for higher galaxy overdensities, we find a suppression of the gas accretion rate for centrals at $z=0$, similar to (but less strong than) the effect of overdensity on satellites. This means that there is some environmental dependence not determined by halo mass, possibly because strongly clustered central galaxies formed earlier, compete for the same gas reservoir, and/or were satellites previously. 

Stellar and AGN feedback can reduce the gas accretion rate onto galaxies by an order of magnitude \citep{Voort2011a, Faucher2011}. This is due to the fact that outflows do not only eject ISM gas from galaxies, but also plough into inflowing gas, thus preventing it from accreting. Although feedback is necessarily modelled with simplified subgrid recipes in EAGLE, the fact that this simulation reproduces the galaxy stellar mass function lends credence to the idea that the implemented subgrid models capture the large-scale effects of the outflows reasonably well. Particularly relavant for our study is that \citet{Marasco2016} have shown that EAGLE reproduces the observed dependence of H\,\textsc{i} content on environment. Even if the effect of feedback is not captured well and galactic winds behave differently in reality, this will primarily affect the normalization of the relations discussed in this work, but the trends with stellar mass, environment, and redshift are likely robust to changes in feedback models. 

The suppression of gas accretion in dense environments has direct bearing on our understanding of the quenching of satellite galaxies if they are indeed (in part) quenched through starvation. There is some observational evidence for long quenching times, which could point to starvation as the quenching mechanism, since galaxies will slowly consume their remaining ISM until they run out of fuel for star formation \citep[e.g.][]{Wetzel2013, Wheeler2014, Peng2015}. The large scatter in gas accretion rates at fixed stellar mass can potentially explain the large range in observed SFRs in cluster environments \citep[e.g.][]{Chies2015, Paccagnella2016}. The smaller effect of the environment on galaxy SFRs observed at higher redshift \citep[e.g.][]{Scoville2013, Lin2016, Kawinwanichakij2016, Nantais2016, Hatfield2016} is consistent with the reduced environmental dependence of the gas accretion rates found in our simulations. Since, for a fixed environment, gas accretion is more strongly suppressed in lower-mass galaxies, we expect that quenching of their star formation proceeds more rapidly, if starvation indeed plays an important role in galaxy quenching. 

The SFR in our simulations follows the gas accretion rate closely. It exhibits the same trends with environment, although it is somewhat less suppressed than the gas accretion rate. This is likely due to the fact that a galaxy is not immediately stripped of its ISM and is replenished by stellar mass loss, so that it can continue to form stars even if its gas supply is completely shut off until it runs out of fuel. This leads us to conclude that, at least for small galaxies in dense environments, the quenching of star formation could directly result from the suppression of gas accretion, which is strongly affected by the properties of the surrounding gas. However, we did not investigate the removal of the ISM through environmental processes, such as ram pressure stripping or tidal stripping, which could dominate the quenching of some or all satellites. In the absence of stripping, outflows, and mergers, most galaxies at $z=2$ are building up their ISM, whereas massive galaxies at $z=0$ are depleting it. More massive galaxies and satellites in denser environments generally deplete their gas reservoirs more quickly, consistent with the observed downsizing of galaxies and the SFR--density relation (see Figure~\ref{fig:sfr}). However, at fixed halo mass, satellites in groups and clusters at $z=0$ deplete their ISM less quickly with increasing stellar mass due to the decreasing environmental dependence of their gas accretion and star formation rates. 

The environmental dependences found in our hydrodynamical simulation could help inform and improve the treatment of gas accretion onto satellites in semi-analytic models \citep[e.g.][]{Guo2016}. We focused on the `instantaneous' gas accretion rates for central and satellite galaxies. As hinted at by the dependence on halocentric radius (see Figure~\ref{fig:rad}), the accretion rate onto a satellite is affected by the amount of time spent in the host halo and/or its trajectory. The timescales over which the gas accretion rates decrease after galaxies fall into massive haloes will be explored in future work (Bah\'e et al.\ in preparation), which will help to understand the origin of the large scatter in accretion rates shown in this paper.

\section*{Acknowledgements}

We would like to thank Lihwai Lin, Tim Davis, Chiara Tonini, and Miguel Arag\'on-Calvo for useful discussions and the Simons Foundation and participants of the Simons Symposium `Galactic Superwinds: Beyond Phenomenology' for inspiration for this work.
We would also like to thank the referee for valuable comments that helped clarify some important points. 
This work used the DiRAC Data Centric system at Durham University, operated by the Institute for Computational Cosmology on behalf of the STFC DiRAC HPC Facility (www.dirac.ac.uk). This equipment was funded by BIS National E-infrastructure capital grant ST/K00042X/1, STFC capital grant ST/H008519/1, and STFC DiRAC Operations grant ST/K003267/1 and Durham University. DiRAC is part of the National E-Infrastructure. 
This work was supported by the Klaus Tschira Foundation, by the Netherlands Organisation for Scientific Research (NWO), through VICI grant 639.043.409, by the European Research Council under the European Union's Seventh Framework Programme (FP7/2007- 2013) / ERC Grant agreement 278594-GasAroundGalaxies, and by the Interuniversity Attraction Poles Programme initiated by the Belgian Science Policy Office (AP P7/08 CHARM).
RAC is a Royal Society University Research Fellow. The simulation data used in this study are available through collaboration with the authors.

\bibliographystyle{mnras}
\bibliography{satellites}

\bsp

\appendix

\section{Resolution test} 

\begin{figure*}
\center
\includegraphics[scale=.52]{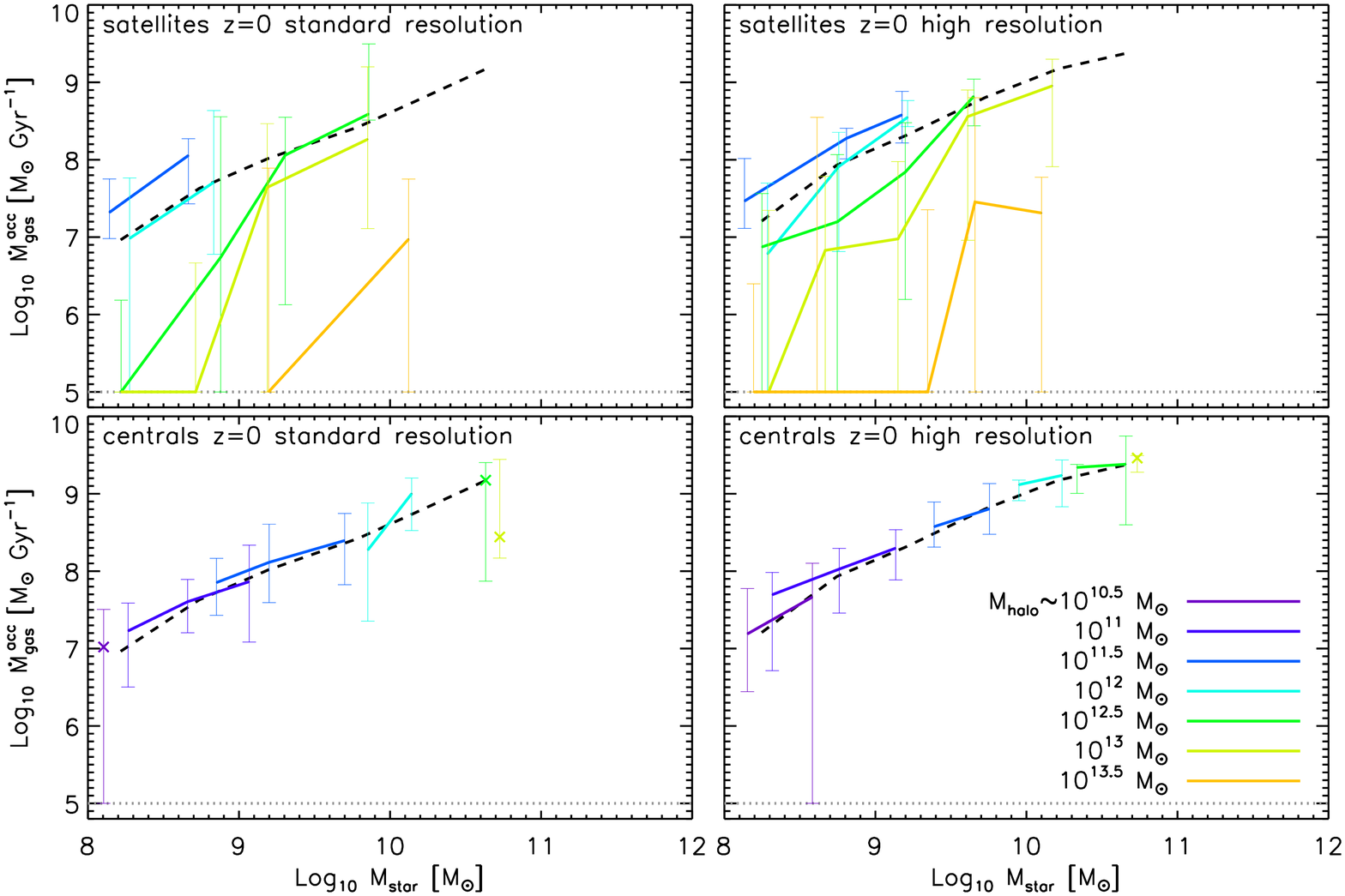}
\caption {\label{fig:res} Median gas accretion rate onto galaxies as a function of stellar mass for a (25~comoving~Mpc)$^3$ simulation at standard resolution (\emph{``Ref-L25N376''}; left) and a (25~comoving~Mpc)$^3$ simulation at 8 times higher mass resolution (\emph{``Ref-L25N752''}; right) for satellites (top) and central galaxies (bottom) at $z=0$. The dashed curves show median accretion rates for all galaxies. Solid, coloured curves and crosses show median accretion rates in halo mass bins, 0.5~dex in width, centred on the values indicated in the legend. Gas accretion rates of zero have been set to $10^5$~M$_\odot$~Gyr$^{-1}$. Error bars show the 16th and 84th percentiles of the distribution. The bins are twice as large in $M_\mathrm{star}$ as in Figure~\ref{fig:mass}, to increase statistics, and only bins containing at least 5 galaxies are shown. The suppression of gas accretion onto satellites due to the environment, visible in the top panels, is independent of resolution.}
\end{figure*}
For our resolution test we use a simulation with the same resolution as the one used for our main results, but with a volume that is smaller by a factor of 64 (\emph{``Ref-L25N376''}). The simulation volume limits the total number of galaxies and which environments we can probe, but there is no volume effect at fixed environment.  We have repeated our analysis at 8 (2) times higher mass (spatial) resolution using a simulation with the same feedback parameters (\emph{``Ref-L25N752''}) and one with recalibrated parameters to again match the stellar mass function and galaxy sizes (\emph{``Recal-L25N752''}). Note that the former simulations do not match these observables as well for numerical reasons \citep{Schaye2015}. We find that the normalization of the gas accretion rates depends on resolution. At fixed resolution, it depends on the (recalibrated) feedback parameters, but only for massive galaxies ($M_\mathrm{star}\gtrsim10^{10}$~M$_\odot$). Higher resolution leads to increased gas densities, which result in higher cooling rates and higher accretion rates. The difference in gas accretion rates between our standard resolution simulation and the higher-resolution simulations is 0.3~dex on average.

Despite this normalization issue, the trends with stellar mass, halo mass, and galaxy overdensity remain the same, as shown in Figure~\ref{fig:res}. This figure shows the median gas accretion rate as a function of steller mass for simulation \emph{``Ref-L25N376''} (left) and \emph{``Ref-L25N752''} (right) at $z=0$. Satellites (centrals) are shown in the top (bottom) panels in different halo mass bins, as shown by the coloured curves. The dashed curves show median accretion rates for all galaxies in each simulation. Gas accretion rates of zero have been set to $10^5$~M$_\odot$~Gyr$^{-1}$ and the error bars show the scatter around the median. The environmental suppression of gas accretion in satellites is insensitive to resolution. We therefore conclude that the behaviour of the gas accretion onto galaxies, described in this paper, is robust. Note, however, that we are unable to probe cluster environments in these small simulation volumes.

\label{lastpage}

\end{document}